\documentclass[letterpaper]{article}

\usepackage[T1]{fontenc}

\usepackage{geometry}
\usepackage{setspace}

\usepackage[style = chem-acs, doi = true, maxnames = 15, articletitle = true]{biblatex}
\usepackage{graphicx}
\usepackage{float}
\newfloat{scheme}{htbp}{los}
\floatname{scheme}{Scheme}
\floatname{chart}{Chart}
\newfloat{graph}{htbp}{loh}

\usepackage{chemformula} 
\usepackage[version = 4]{mhchem} 
\usepackage{hyperref}
\usepackage{amssymb}

\newcommand{\fcunit}{eV/\AA$^2$\ }

\usepackage{authblk}
\author[1]{Mohammad Bagheri*}
\author[1]{Pekka Koskinen*}
\affil[1]{Nanoscience Center, Department of Physics, University of Jyv\"askyl\"a, 40014 Jyv\"askyl\"a, Finland}

\title{Machine-Learning-Accelerated Metallene Stabilization\\ from High-Throughput Sandwich Modeling}
\date{*Email: mohammad.m.bagheri@jyu.fi, pekka.j.koskinen@jyu.fi}

\begin{document}

\maketitle

\begin{abstract}
 Metallenes have appealing properties, but stabilizing them in a monolayer phase poses synthesis challenges. A recent experiment showed that the van der Waals squeezing method can stabilize certain metallenes in a \ce{MoS2} sandwich. This pioneering work motivates systematic studies, but such studies are experimentally impractical, while first-principles modeling remains prohibitive. Here, using universal machine-learning interatomic potentials, we constructed $1620$ metallene sandwich heterostructures containing $6$ different sandwich layers and $45$ metals. We performed phonon calculations, which revealed $1208$ dynamically stable structures and narrowed them to $331$ after structural filtering. We found that Se-based transition-metal dichalcogenides stabilize metallenes effectively. Specifically, hexagonal and buckled hexagonal crystal lattices exhibit the greatest stability. We further evaluated the thermal stability of selected heterostructures with density-functional theory molecular dynamics simulations at room temperature. By uncovering the physical and chemical factors that govern metallenes stabilization, our results provide systematic insights to guide and accelerate synthesis for future applications.
 
 \vspace{0.8em}
\end{abstract}

Metallenes are atomically thin, two-dimensional (2D) elemental metals with unique properties that are highly promising for diverse technological applications \cite{Jiang_2023, Wenxuan_2024, Lu2023_bio, shahzad2024recent, Xie2023, yaoda2020, Hongyu_2024, Kashiwaya_2025, Kabiraz_2025, Wenbo_2026, Du2026}. Computational studies \cite{atlas, Ono_2020, gentle, Yang_2026} have demonstrated that isolated metallenes, particularly transition metals, can achieve dynamical stability. However, in practice, their isotropic metallic bonding makes it challenging to synthesize free-standing forms. Consequently, developing effective stabilization strategies remains essential to advancing their practical applications.

While to date numerous strategies have been developed for synthesizing metallenes \cite{goldene2024, Ren2025}, practical experimental approaches relied until recently on growing metallenes in 2D materials pores or confining them within 2D templates \cite{Zr_patch, Zhao_Mo, Zhao_2D_gold, Zhu_2D_gold, Ta_Cr, Mendes_Zr, Zhao_2025}.
Recent experimental progress has demonstrated a van der Waals (vdW) squeezing technique that stabilizes 2D metals (including Bi, Ga, In, Sn, and Pb) by encapsulating them between two \ce{MoS2} monolayers \cite{Zhao_2025}. This strategy addresses the long-standing challenge of 3D clustering in non-vdW materials, including metallenes, by exploiting the structural confinement principles inherent to stacked vdW heterostructures, thereby facilitating more applied experimental research on metallenes \cite{Geim2013, Cui_2026, Cui_2026_NL}. 

Building upon this experimental framework, subsequent computational research has investigated the behavior of 2D metals within h-BN sandwiches, identifying several additional stable heterostructures \cite{Zhang_2025}; however, a systematic exploration of a broader library of sandwiching layers and an expanded inventory of candidate metallenes is still lacking. This expansion would map out the vast, uncharted landscape of interfacial phases and predict stable configurations.

For systematic exploration, first-principles calculations can provide highly valuable insights, but simulating a large number of heterostructures with large supercells is computationally costly and impractical. To overcome this obstacle, universal machine-learning interatomic potentials (UMLIPs) offer a powerful alternative; they have advanced to deliver first-principles-level accuracy and possess exceptional scalability, enabling accelerated studies at a fraction of the computational expense \cite{alexandria, benedini2025, Kristian2025, Burger2025, interface, Han_2025, MLFCDimen}.

Therefore, in this Letter, we used a computational approach to design and validate the existence of metallene sandwich heterostructures (MSH) using UMLIP.
We evaluated our workflow against first-principles calculations for several representative cases and confirmed its accuracy.
We constructed $1620$ MSHs using h-BN, four transition-metal dichalcogenides (TMDs), and graphene as sandwich layers, with monolayers of $45$ elemental metals in $6$ different lattices (honeycomb, square, hexagonal, and their buckled forms).
We optimized the MSHs, performed phonon calculations, and identified $1208$ dynamically stable MSHs, which we narrowed to $331$ after structural analysis. We identified TMDs as the most effective sandwich layer for stabilizing metallenes. Overall, metals with hexagonal (hex) and buckled hexagonal (bhex) lattices were stabilized the most efficiently. In addition, we evaluated the thermal stability of selected MSHs with density-functional theory (DFT) molecular dynamics (MD) simulations at room temperature.
Furthermore, to understand the stabilization mechanisms of sandwiched metallenes, we analyzed sandwich-induced biaxial strains, interlayer force constants (FCs) between the supporting and metal layers, partial charge densities, and charge transfer through charge density difference and Bader charge analyses.
Finally, we investigated the electronic properties of selected MSHs by calculating band structures and projected density of states (PDOS).

\begin{figure*}
\centering
  \includegraphics[width=\textwidth]{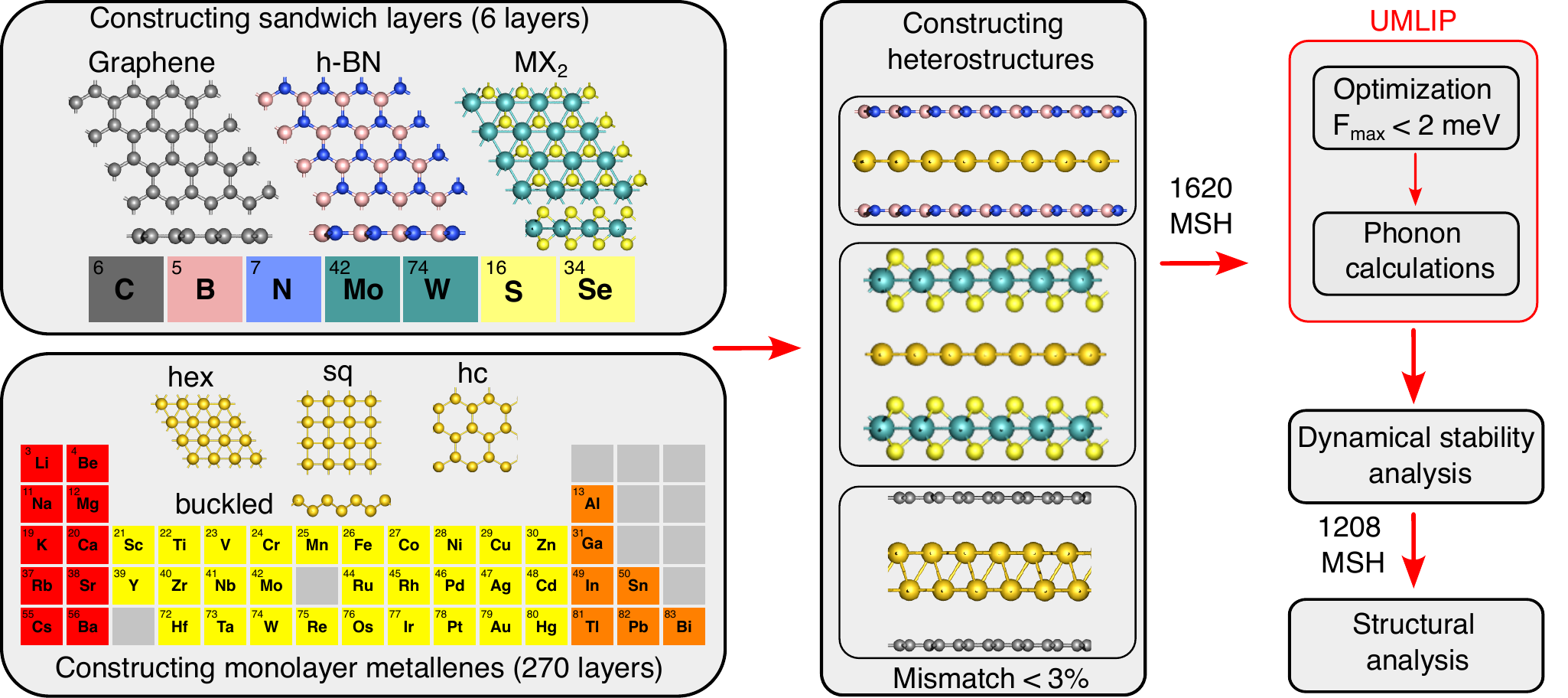}
\caption{\label{fig:heterostructure}
 \textbf{High-throughput calculation workflow for metallene sandwich heterostructures (MSHs)}. Sandwich layers consist of graphene (C), hexagonal boron nitride (h-BN), and TMDs with the chemical formula of \ce{MX2}, where M: Mo, W, and X: S, Se). The atomic species colors are indicated in insets. Metallene with three planar lattices: hexagonal (hex), square (sq), and honeycomb (hc), and their buckled forms (bhex, bsq, and bhc). The list of metals used is shown in the periodic table. Schematic illustrations of MSHs for three examples are shown. The universal machine-learning interatomic potentials (UMLIP) stage is indicated with a red box.
}
\end{figure*}

In our high-throughput calculations workflow (Figure~\ref{fig:heterostructure}), we first designed the MSH, which consists of a monolayer of metallene and two sandwich monolayers, and then performed relaxation and phonon calculations.
We constructed monolayer metallenes in three planar lattices: hexagonal (hex), square (sq), and honeycomb (hc), with their buckled forms (bhex, bsq, and bhc), for a total of six different metallene layers. For the sandwich layers, we selected six 2D materials: graphene, h-BN, and four TMDs (\ce{MoS2}, \ce{MoSe2}, \ce{WS2}, and \ce{WSe2}). 
To construct the initial heterostructure, we created supercells of metallenes and sandwich layers, ensuring that the lattice mismatches between supercells remained below $3$\% (for structural compatibility) by expanding the supercell until the difference in cell sizes fell below the targeted mismatch. 
Overall, $1620$ MSHs were constructed. The number of atoms in every MSH is found in Figures~S1--S3. We use S|M(L)|S as a general notation for MSHs, where S is the sandwich layer, M is the metallenes, and L is the metallene's lattice.
Next, we optimized MSHs, selecting only structures that were fully converged and met the force criteria for phonon calculations. After performing phonon calculations, the dynamical stabilities of MSHs were analyzed, and unstable materials (imaginary frequencies $>0.2$~THz) were removed. The threshold ($0.2$~THz) for imaginary frequencies, chosen based on our benchmarks, previous works, and literature \cite{alexandria, MatterSim, Burger2025, interface, MLFCDimen}, is a reasonable threshold considering the possible shifts in low frequencies from UMLIPs. 
To ensure structural reliability, we evaluated final trajectories for severe structural distortions, checking lattice and interfacial conditions. Structures were flagged as unstable and filtered out if interatomic distances dropped significantly compared to the sum of their covalent radii, initial layer stacking order got inverted, individual layer thickness expanded and large bending occurred (root-mean-square greater than $1$\AA), phases became mixed (for example hexagonal and honeycomb), or the metallene transformed into another prototype.
We note that the high-throughput phonon screening was performed with MatterSim, whereas the subsequent structural analysis and binding energies included the D3 dispersion correction (see SI for details).

\begin{figure*}
\centering
  \includegraphics[width=\textwidth]{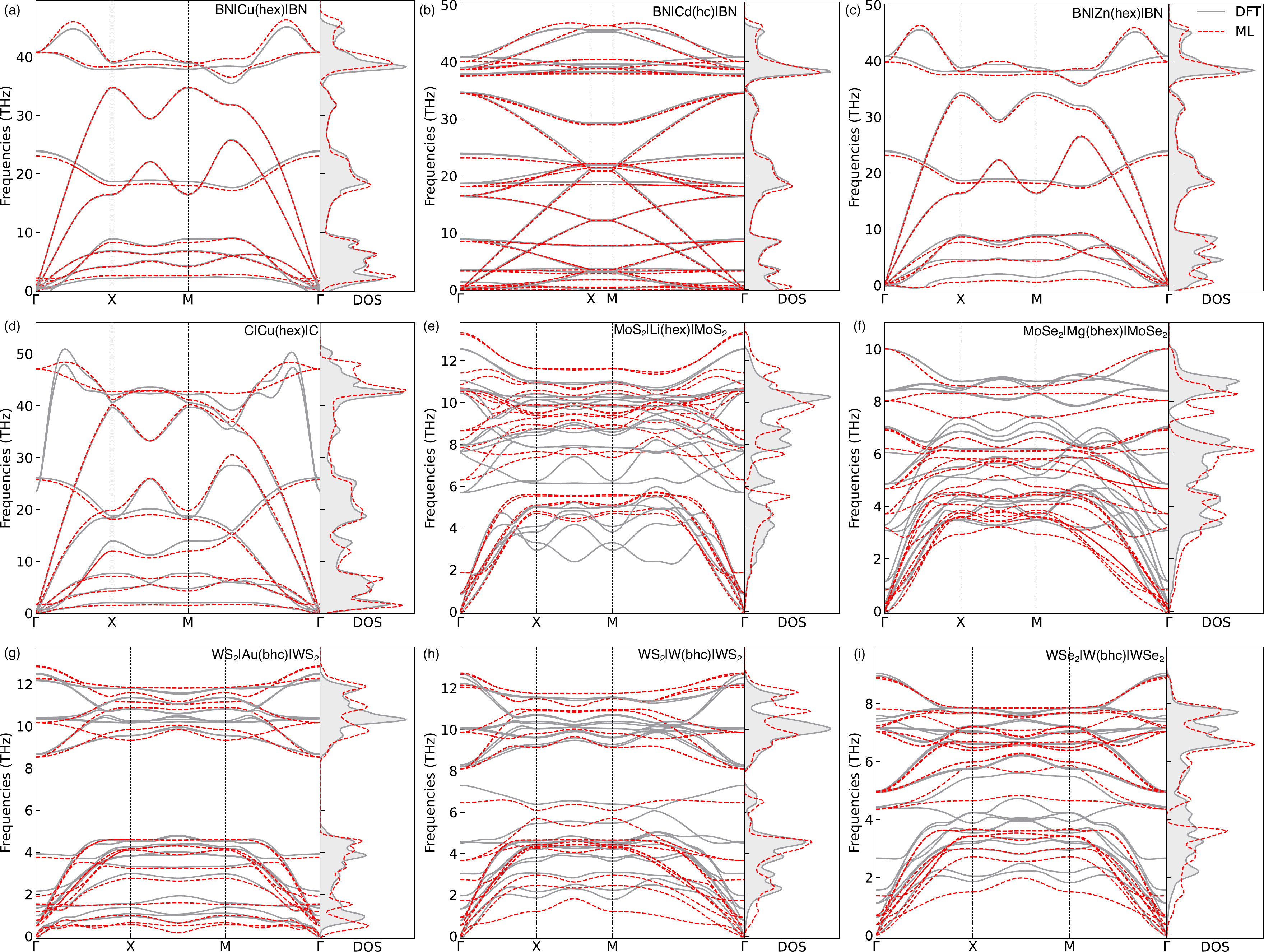}
\caption{\label{fig:benchmark}
 \textbf{Comparison of first-principles and UMLIP calculations for selected MSHs}. Phonon dispersion curves and phonon density of states of (a) BN|Cu(hex)|BN, (b) BN|Cd(hc)|BN, (c) BN|Zn(hex)|BN, (d) C|Cu(hex)|C, (e) \ce{MoS2}|Li(hex)|\ce{MoS2}, (f) \ce{MoSe2}|Mg(bhex)|\ce{MoSe2}, (g) \ce{WS2}|Au(bhc)|\ce{WS2}, (h) \ce{WS2}|W(bhc)|\ce{WS2}, and (i) \ce{WSe2}|W(bhc)|\ce{WSe2}. The DFT bands are shown as solid gray lines, and the UMLIP bands as dashed red lines.
}
\end{figure*}

Here, we chose MatterSim as the primary UMLIP calculator in our workflow, as it shows excellent accuracy in materials properties, including energies, forces, FCs, phonons, and defects \cite{alexandria, benedini2025, Kristian2025, Burger2025, Shuang_2025, interface, Anam2025, Han_2025, MLFCDimen, Sharma2025}.
Before proceeding further, we evaluated and validated the UMLIP results by performing the same workflow and using first-principles calculations for eleven reasonably sized MSHs, BN|Cu(hex)|BN, BN|Cd(hc)|BN, BN|Zn(hex)|BN, C|Cu(hex)|C, \ce{MoS2}|Li(hex)|\ce{MoS2}, \ce{MoSe2}|Mg(bhex)|\ce{MoSe2}, \ce{WS2}|Au(bhc)|\ce{WS2}, \ce{WS2}|W\allowbreak(bhc)\allowbreak|\allowbreak\ce{WS2}, and \ce{WSe2}|W(bhc)|\ce{WSe2}, C|Zn(sq)|C, and \ce{MoS2}|Mg(hex)|\ce{MoS2}.

We benchmarked various configurations, including all the sandwich layers, several metal lattices (hex, hc, sq, and buckled ones), and metals from all over the periodic table, including alkali metals (Li), alkaline-earth metals (Mg), transition metals (W, Cu, Au), and late transition metals (Cd, Zn). We have to note that most MSHs are not practically possible to be calculated using DFT due to the large number of atoms in their structures.
Since the usability of MatterSim has been extensively evaluated, and the details are in the literature, we focus only on comparing the geometrical properties (bond lengths and interlayer distance), phonon curves, and phonon density of states.
We used DFT to calculate reference structural properties, yielding results that were close to UMLIP's in both cases for average bond lengths and interlayer distances (Table~S1).
Next, we used DFT to calculate phonon dispersions and phonon density of states. The selected structure demonstrated in Figure~\ref{fig:benchmark} has no imaginary frequencies ($>0.2$~THz), and the UMLIP-calculated phonons are pretty close to DFT, with only small shifts in higher frequency bands for MSHs with graphene and h-BN as sandwich layers. The shifts in phonon bands for TMD cases are larger, but the stability results, which are the main focus here, are accurate. The phonon density of states behaves pretty similarly to the phonon bands.
We illustrate C|Zn(sq)|C, and \ce{MoS2}|Mg(hex)|\ce{MoS2} with larger imaginary frequencies (Figure~S4) as two unstable examples.
In \ce{MoS2}|Mg(hex)|\ce{MoS2}, DFT resulted in imaginary frequencies larger than our threshold, which are comparable to UMLIP phonons. Although shifts in the phonon bands are larger than in the C|Zn(sq)|C system, the UMLIP prediction still serves our purpose of distinguishing dynamically unstable MSHs.

For the low-frequency phonons, we compared MatterSim and DFT for the five TMD-based systems with the largest deviations and for the unstable C|Zn(sq)|C (Figure~S5). Both methods yield the same stability classification for all systems. Where MatterSim deviates for the softest modes, it mostly underestimates the DFT frequencies, making the screening conservative. This agrees with large-scale benchmarks showing that MatterSim reliably distinguishes dynamically stable from unstable materials~\cite{alexandria}. Studies that require quantitatively accurate phonon spectra may benefit from fine-tuning the model on interface-specific DFT data.

After validating the DFT-level accuracy of MatterSim, we performed high-throughput UMLIP-based calculations to investigate the dynamical stability of all constructed MSHs and analyze their structures (Figure~\ref{fig:heterostructure}).
The dynamical stability analysis identified $1208$ stable MSHs (Figure~S6), which were narrowed to $331$ final MSHs after structural analysis.
The excluded structures are summarized in Table~S2. However, most dynamically stable structures that didn't pass our structural filtering may exist in practice, especially those with mixed phases of hex with bhex, and hc with hex.

\begin{figure*}[t]
\centering
  \includegraphics[width=0.8\textwidth]{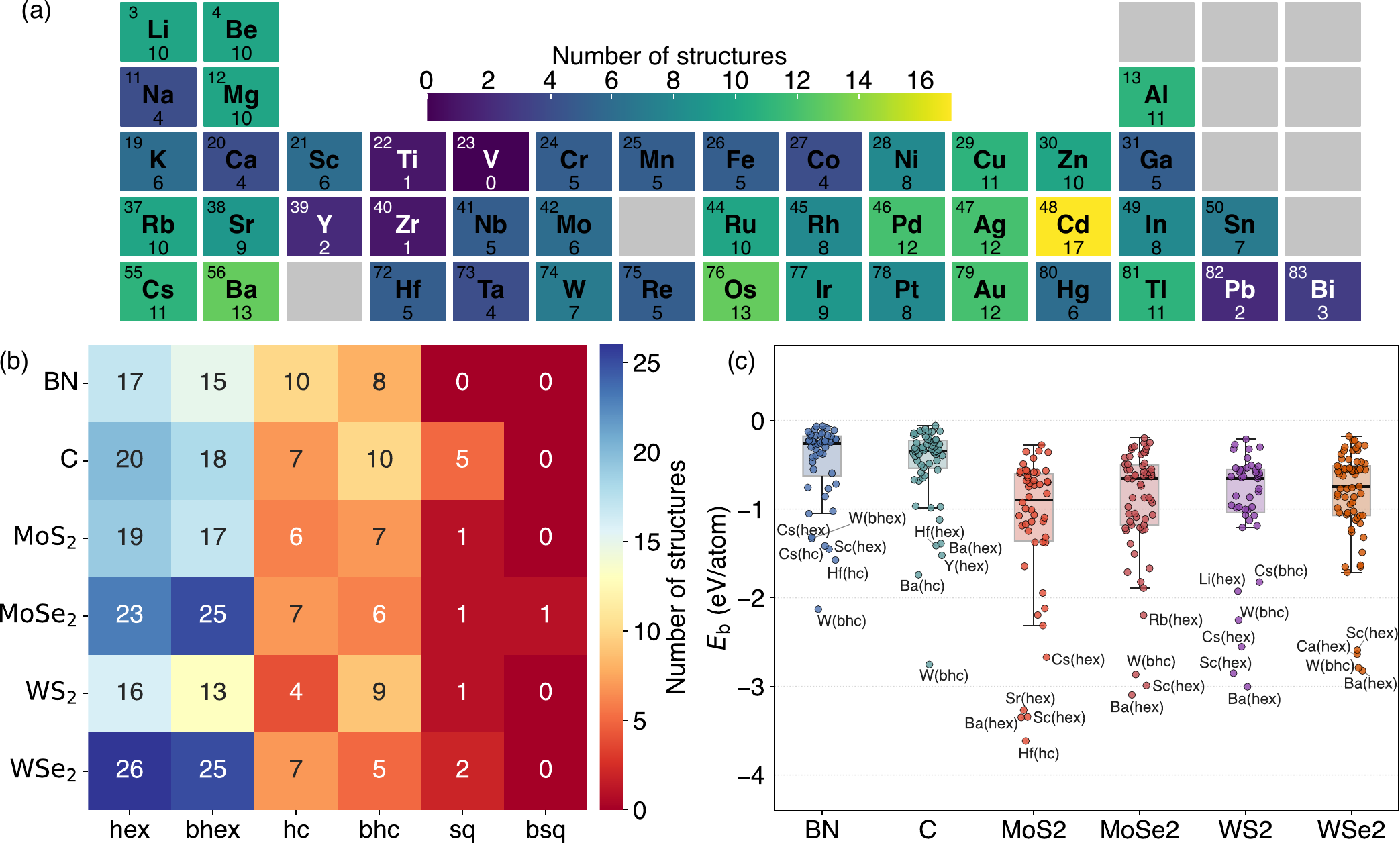}
\caption{\label{fig:dyn}
  \textbf{Dynamical stability and structural analysis.} (a) Heatmap of the total number of dynamically stable (no imaginary frequency $>0.2$ THz) and structurally filtered MSHs. (b) Number of dynamically stable and structurally filtered sandwich heterostructures vs. sandwich layer type, grouped by metallenes' lattice. (c) Calculated binding energy ($E_{\mathrm{b}}$) of the stable MSHs grouped by the sandwich layer. Outlier metals and their lattices are indicated.
}
\end{figure*}

The phonon and structural analysis of final MSHs reveal that the investigated sandwich-layer configurations consistently stabilize the metallene monolayers, though their stabilizing impact varies by configuration (Figure~\ref{fig:dyn}a,b).
The alkali and alkaline Earth metals (groups $1$ and $2$) generally show high stability counts (e.g., Ba with $13$, Cs with $11$, and Li, Be, Mg, Rb with $10$), with Na ($4$), Ca ($4$), and K ($6$) having the fewest stable structures in these groups.
In transition metals, early transition metals have very low stability (notably V with no stable structure, Ti and Zr with $1$, and Y with $2$), whereas late transition metals show significantly higher numbers of stable structures, with Cd show the highest overall count ($17$), followed by Os ($13$), Pd ($12$), and Au ($12$).
Post-transition metals show varied stability, ranging from high counts in Al ($11$) and Tl ($11$) to very low counts in Pb ($2$) and Bi ($3$).

The number of stable MSHs differs considerably among the sandwich layers. \ce{WSe2} has $65$, \ce{MoSe2} has $63$, graphene has $60$, h-BN and \ce{MoS2} have $50$, and \ce{WS2} $43$ stable MSHs out of $270$ possible candidates (Figure~\ref{fig:dyn}b). Thus, the Se-based TMDs stabilize the largest number of metallenes, whereas the S-based TMDs support numbers comparable to or lower than h-BN. This difference arises mainly from buckled hexagonal metallenes, $25$ of which are stable in each Se-based TMD compared with $17$ in \ce{MoS2} and $13$ in \ce{WS2}. h-BN is the only sandwich layer with no stable MSHs with the sq lattice. However, the number of stable MSHs alone does not reflect how strongly a sandwich layer stabilizes a given metallene, which we analyze further in terms of binding energies.

In addition, we found that metallenes with hex and bhex lattices stabilize most effectively, whereas those with bsq (only one MSH) and sq lattices stabilize the least effectively. The least stable MSHs belong to S|M(bsq)|S, followed by BN|M(sq)|BN (see Figures~S7--S23 for details).
The first experimental study \cite{Zhao_2025} characterized several 2D metals, including Bi, Ga, In, Sn, and Pb, between two \ce{MoS2} monolayers and further used DFT to investigate the properties of Bi alone, thereby supplementing experimental findings. Our findings identify stable Bi phases across various sandwich structures. These phases predominantly appear in honeycomb (hc) and buckled (bhc) forms, consistent with experimental observations (Figure~S24).
We found that hc is the most stable lattice for Ga and Sn (appearing on all $6$ sandwich layers), In favors the bhex configuration, and Pb shows more stable MSHs with the bhex lattice.
While alkali, alkaline earth, and transition metals remain stable across most sandwich layers, selecting a specific layer is critical for early- and post-transition metals, where many combinations fail to form stable structures. We also found that the sandwich layer directly influences the monolayer's structural configuration, as specific lattices can only exist when paired with certain layers.

We have also calculated the binding energy ($E_{\mathrm{b}}$) of the final stable MSHs (see SI for more details).
All stable MSHs possess negative binding energies (Figure~\ref{fig:dyn}c), indicating favorable interfacial interaction relative to the isolated layers.
Sandwiching metals in h-BN and graphene is governed primarily by vdW interactions, with median binding energies of $-0.26$ and $-0.34$~eV/atom, respectively. In contrast, TMD sandwiches show increased binding, with medians shifting to $-0.89$~eV/atom for \ce{MoS2}, $-0.74$~eV/atom for \ce{WSe2}, and $-0.65$~eV/atom for both \ce{MoSe2} and \ce{WS2}. While most metallenes in h-BN and graphene experience gentle mechanical confinement without severe electronic perturbation, prominent negative outliers are observed for hex, hc, and their buckled forms. The early transition metals and alkali/alkaline-earth metals, most notably W(bhc) (down to $-2.75$~eV/atom in graphene and $-2.13$~eV/atom in h-BN), as well as Ba, Hf, Sc, and Cs, show possible hybridization or charge transfer. In TMDs, this chemical interaction strengthens even further and outliers such as Hf(hc) in \ce{MoS2} ($-3.62$~eV/atom), Ba(hex) ($-2.82$ to $-3.35$~eV/atom), Sc(hex) ($-2.59$ to $-3.34$~eV/atom), and W(bhc) (up to $-2.87$~eV/atom) highlight strong interfacial interactions.
Since these medians include different sets of metallenes for each sandwich layer, we further compared the same metallene and lattice across sandwich layers (Table~S3). Each TMD binds more strongly than graphene in $96-100$\% of the cases and more strongly than h-BN in all cases. Among the TMDs, S-based TMDs interact most strongly with a given metallene, whereas Se-based TMDs stabilize more metallenes.
A negative binding energy, however, does not by itself establish stability against competing configurations. We therefore compared each MSH with the bulk metal plus the isolated sandwich layer (Figure~S25). As for free-standing metallenes, the bulk metal remains lower in energy for most MSHs. However, encapsulation lowers the energy relative to the free-standing metallene for $77-96$\% of the TMD-based MSHs, compared with $63$\% for graphene and $48$\% for h-BN. Encapsulation thus dynamically stabilizes metastable metallenes and, particularly for TMDs, also lowers their energy, consistent with experimental routes to 2D metals that yield metastable phases protected by encapsulating layers~\cite{Zhao_2025}.

\begin{figure*}
\centering
  \includegraphics[width=\textwidth]{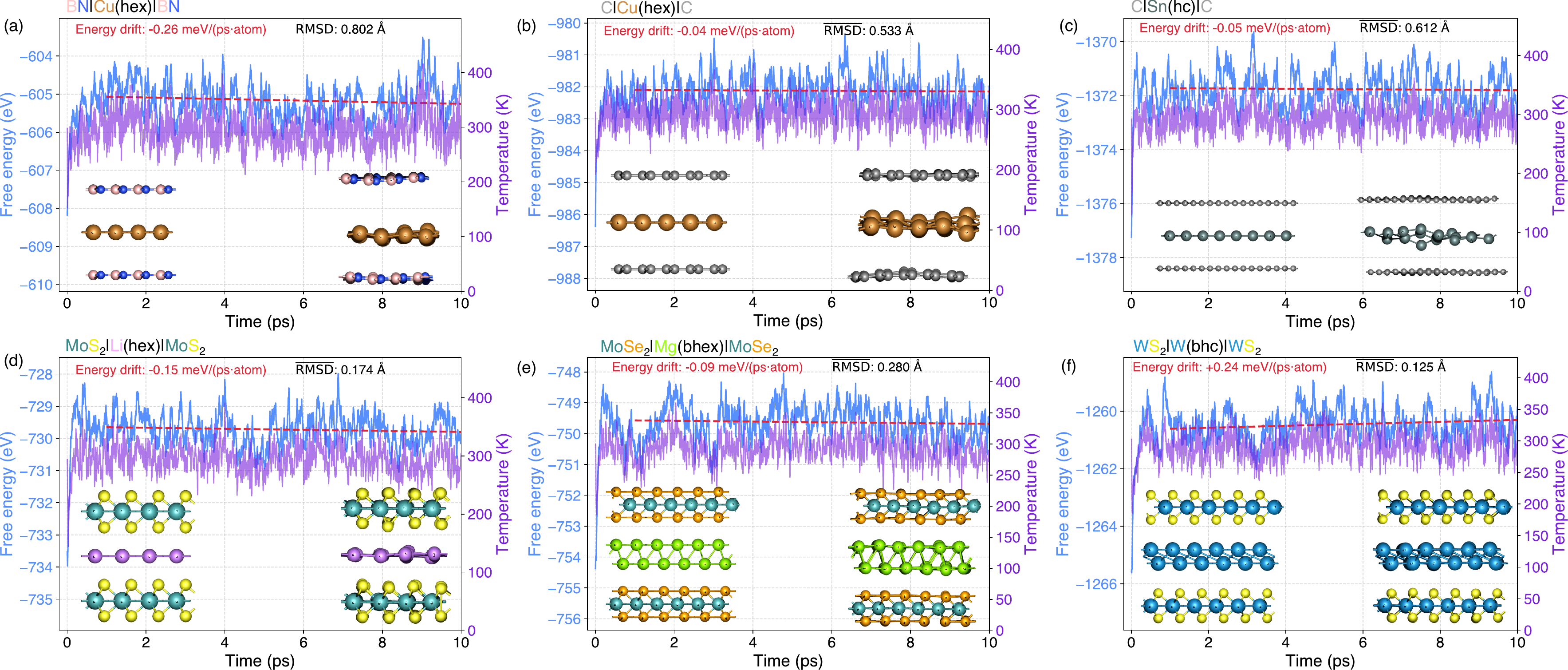}
\caption{\label{fig:MD}
 \textbf{Finite-temperature structural stability.} DFT MD simulations of selected stable MSHs at room temperature for $10$~ps. The blue line indicates free energy, and the purple line indicates temperature evolution in the MD simulation. The insets show the side views of the initial ($0$~ps) and final ($10$~ps) structures. The time-averaged root-mean-square displacement ($\overline{\text{RMSD}}$) and the normalized energy drift rate per atom are indicated ($t \ge 1\text{ ps}$). The red dashed line represents the linear regression fit to the free energy over the MD.
}
\end{figure*}

To evaluate the finite-temperature structural stability of MSHs, we performed DFT-based MD simulations of $6$ selected dynamically stable MSHs (see Table~S4 for more details).
The free energy remained within a limited range of fluctuations and energy drift throughout the MD simulations, and despite minor thermal rippling, no structural reconstruction, phase transition, or bond breaking was observed, confirming that these configurations maintain their structural integrity under ambient conditions (Figure~\ref{fig:MD}). In all the selected heterostructures, the time-averaged root-mean-square displacement ($\overline{\text{RMSD}}$) remained less than $1$\AA, confirming our choice in structural analysis.  Bond-length (metal-metal) analysis in MD simulations, which governs thermal fluctuations and the structural stability of the metallic layer at finite temperature, also confirmed the structural stability (see Figure~S26 and SI for more details).
We have to note that $10$~ps MD simulations confirm kinetic and thermal stability against spontaneous room-temperature structural reconstruction or disintegration, rather than global thermodynamic stability.

To discuss the stabilization mechanism of metallene sandwiches, let us first recall the inherent structural limitations of these systems. Metallenes are fundamentally defined as yielding membranes that lack the inherent structural stiffness required to remain flat at their natural energy minima.
We investigated the stabilization mechanisms by analyzing sandwich-induced metallene strains ($\varepsilon$, tensile or compressive) and interlayer FCs.

\begin{figure*}[t]
\centering
  \includegraphics[width=\textwidth]{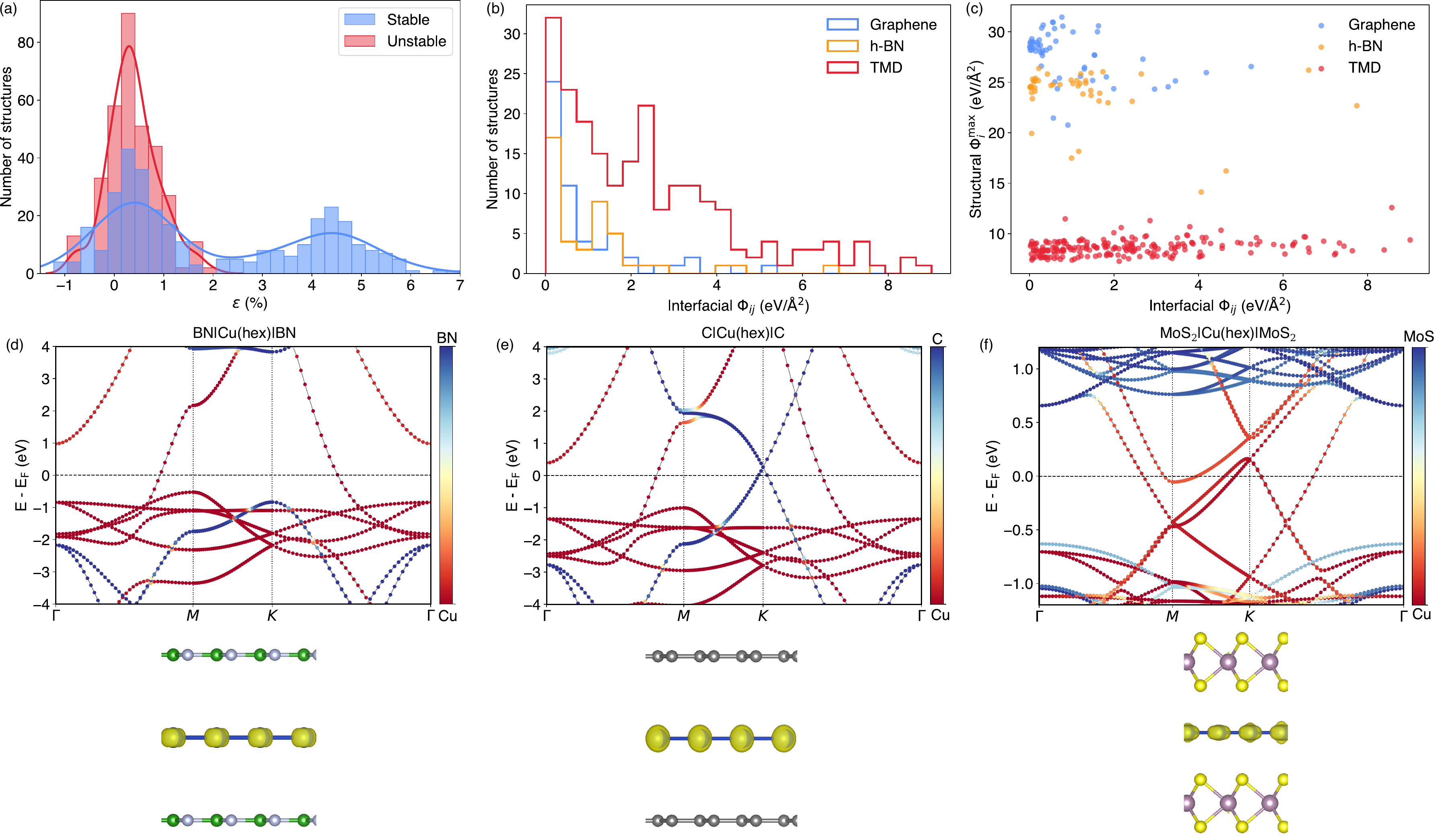}
\caption{\label{fig:mechanism}
  \textbf{Stabilization mechanism of metallene sandwich heterostructures.} (a) Distribution of in-plane strain after relaxation, relative to the initial supercell, for stable (blue) and unstable (red) heterostructures. (b) Number of final stable structures as a function of the interfacial maximum FCs between the metal and the supporting atom from the sandwich layers. (c) Comparison of the structural maximum FCs and interfacial maximum FC between metallene and the supporting atom from the sandwich layers. (d--f) Projected electronic band structures of $3$ sandwich layer representatives: BN|Cu(hex)|BN, C|Cu(hex)|C, and \ce{MoS2}|Cu(hex)|\ce{MoS2} (as TMD representatives), color-coded by contribution of metal and sandwich layers. The partial charge density diagrams near the Fermi level are indicated under each panel.
}
\end{figure*}

Our results demonstrated that the vast majority of metallene layers are stabilized under a small tensile strain, and heterostructures with graphene, h-BN, \ce{MoS2}, and \ce{WS2} exhibit a narrow distribution of strains tightly centered around a minimal tensile strain below $2$\% (Figure~\ref{fig:mechanism}a). This finding aligns well with a previous study showing that gentle tension can stabilize metallenes \cite{gentle, Joudi2025}, as provided here by sandwiching. The \ce{MoSe2}- and \ce{WSe2}-based heterostructures form a second group at $4-5$\%, which arises mainly from the expansion of the supercell toward the larger equilibrium lattice constant of the Se-based TMDs during relaxation, independent of the encapsulated metal (see SI for details). The energy cost of these strains is small: the deformation energy of the metallene is similar for all sandwich layers, with medians of $0.04-0.05$~eV/metal atom, and does not change the stronger binding of TMDs relative to graphene and h-BN (Table~S5). 
We also find that a notable minority of the structures ($41$ out of $331$) stabilize under compressive strain (see Figure~S27 for more details).
In contrast, unstable structures exhibited smaller strains than stable ones within the same sandwich layer (median $0.2$\% versus $0.4$\%, excluding the Se-based TMDs), suggesting that insufficient tension may contribute to their instability.

In addition to biaxial strain, another viewpoint on stabilization is provided by the FCs between the sandwich layer and the metallene atoms. 
Here, we use the \emph{FCDimen} package \cite{FCDimen} approach to calculate FCs (see SI for details), but only considering the interfacial $\Phi_{ij}$ between the metal and sandwich layers.
Focusing on the interfacial $\Phi_{ij}$ between the metal atom and the supporting atom from the sandwich layer (C, B/N, S, and Se) demonstrates that the FC in stable materials is in the range of $1-6$~\fcunit  (Figure~\ref{fig:mechanism}b).
The structural $\Phi_{i}^{\rm max}$ for h-BN and graphene are much larger than the interfacial $\Phi_{ij}$, indicating vdW interactions. In TMDs, the structural FCs are close to the interfacial FCs (Figure~\ref{fig:mechanism}c), suggesting a stronger interaction between metals and supporting atoms. Since FCs quantify the interfacial coupling strength but not its electronic origin, we examine it further using projected band structures, charge density difference, and Bader charge analysis.
We further compared the FC on the metal atom (M) and the support atoms in the layer (Se or C) using \ce{MoSe2} (as a representative TMD) and graphene metallene (hc) heterostructures. For cases that were stable in \ce{MoSe2} but not in graphene, we found that the FCs between M-C are significantly different. Specifically, for Be, Co, Ni, and Pd, the FCs are very small and close to zero (Figure~S28). The FC analysis shows that metallene stabilization also requires sufficient interaction with the sandwich layers, which makes TMDs more effective.

To isolate the role of the sandwich layer, we compare the same metallene and lattice, Cu(hex), sandwiched between h-BN, graphene, and \ce{MoS2}.
We calculated projected band structures and partial charge-transfer distributions near the Fermi level for representatives from each sandwich layer (Figure~\ref{fig:mechanism}d--f). 
In h-BN and graphene, as expected from the FCs analysis, the Cu bands and the sandwich-layer bands near the Fermi level show little mixing; the partial charge density near the Fermi level is localized on the metallic layer, which is consistent with a vdW interaction between the metal layer and the sandwich layers.
In contrast, in \ce{MoS2}, the states at the Fermi level remain mainly of Cu character with a weak \ce{MoS2} contribution, while the \ce{MoS2} valence bands mix markedly with Cu states, indicating hybridization and possible charge transfer.
Since Cu interacts relatively weakly with the TMDs, we additionally examined \ce{WS2}|W(bhc)|\ce{WS2}, one of the strongly bound outliers in Figure~\ref{fig:dyn}c, as an example of the strong-interaction limit (Figure~S29).
We also performed Bader charge analysis to quantitatively calculate charge transfer between sandwich-layer atoms and metallene atoms and validate this finding (Table~S6). The net charge transfer between the metal layer and the graphene or h-BN layers is negligible ($\Delta q \lesssim 0.01\,e$ per metal atom), indicating vdW interactions, while for the same Cu metallene in \ce{MoS2} the charge transfer is much larger, and it becomes substantial for more electropositive metals.
Charge density difference analysis, $\Delta\rho = \rho_{\rm MSH} - \rho_{\rm metal} - \rho_{\rm sandwiches}$, supports this finding (Figure~S30). For h-BN and graphene, the charge redistribution is weak and confined to the vdW gap. For \ce{MoS2}, the redistribution is about seven times larger, with charge accumulating in the Cu--S bonding region at a shorter interlayer distance than in h-BN and graphene, characteristic of Cu--S hybridization; it is stronger still for \ce{WS2}|W(bhc)|\ce{WS2}.

Overall, when metallenes are integrated into a sandwich heterostructure, the monolayer 2D materials act as a physical and chemical scaffold. The sandwich layers can provide tensile strain and FC due to their high Young's moduli. Additionally, the vdW interactions in the case of h-BN and graphene, and hybridization and charge transfer in TMDs as sandwich layers, provide chemical support.
This shifts metallenes away from their unstable equilibrium toward a dynamically stable crystalline state and controls yield behavior through a combination of mechanical pressure and chemical passivation. However, the efficacy of this stabilization is significantly influenced by the choice of the sandwiching 2D material and the intrinsic symmetry of the metallene.

\begin{figure*}
\centering
  \includegraphics[width=\textwidth]{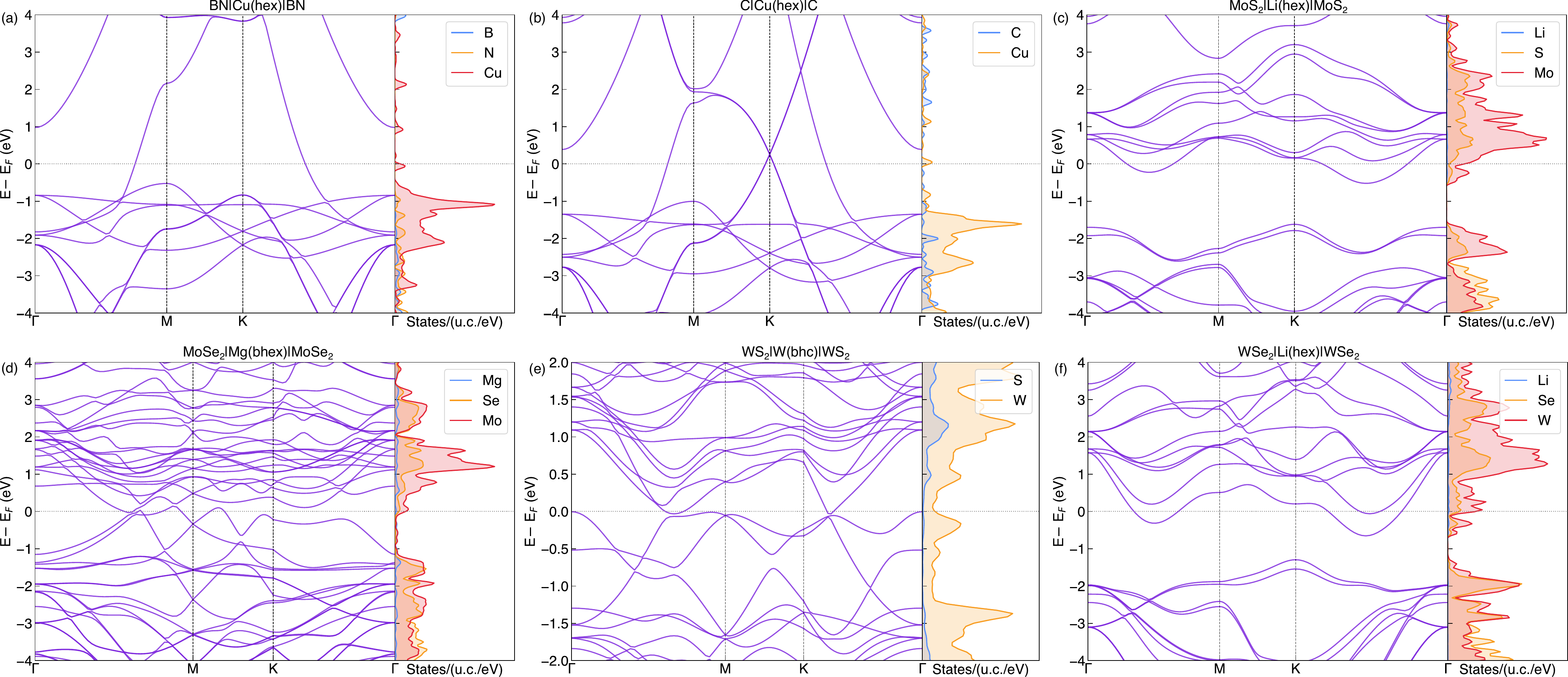}
\caption{\label{fig:ebands}
  \textbf{Electronic structure analysis.} DFT electronic band structures and PDOS of selected MSHs: (a) BN|Cu(hex)|BN, (b) C|Cu(hex)|C, (c) \ce{MoS2}|Li(hex)|\ce{MoS2}, (d) \ce{MoSe2}|Mg(bhex)|\ce{MoSe2}, (e) \ce{WS2}|W(bhc)|\ce{WS2}, and (f) \ce{WSe2}|Li(hex)|\ce{WSe2}.
}
\end{figure*}

To assess whether the encapsulated metallenes retain metallic character within the sandwich architecture, we performed DFT calculations of electronic band structures and PDOS for six representative MSHs (Figure~\ref{fig:ebands}). Across all examined MSHs spanning different sandwich layers, the band structures demonstrate multiple bands crossing the Fermi level. The PDOS reveals a finite electronic density of states at $E_\mathrm{F}$, dominated either by the metallic layer ( in graphene and h-BN MSHs) or mixed metal-sandwich-layer frontier orbitals (in TMD MSHs). While performing electronic structure calculations across the entire dataset lies beyond the scope of the current out-of-the-box UMLIP (without fine-tuning), our previous study of isolated metallenes demonstrated that all candidates except Hg are metallic at the DFT-HSE level~\cite{fermi}. These representative cases suggest that the resulting MSHs preserve metallic behavior in the heterostructure environment, with only Hg-based systems potentially retaining or opening a band gap in some MSHs, which requires further investigation.

In conclusion, we used a universal machine-learning interatomic potential (UMLIP) to systematically study $1620$ metallene sandwich heterostructures with $6$ different sandwich layers (graphene, h-BN, and $4$ TMDs) and various profiles.
We examined our workflow against first-principles calculations for selected representative cases by performing DFT phonon calculations and molecular dynamics simulations.
Our findings confirmed the universality of the sandwiching approach for dynamically stabilizing 2D metals, but the sandwiching layer could influence the formation of monolayer metallene within the heterostructure, particularly the lattice type.
Se-based TMDs support the largest number of stable heterostructures, mainly with hex and bhex lattices, whereas h-BN and graphene are comparatively effective for honeycomb lattices. For the same metallene, however, all TMDs bind more strongly than graphene and h-BN and lower the energy of the metallene more effectively relative to its free-standing form, although the bulk metal remains lower in energy for most heterostructures.
By providing both the mechanical tension to prevent structural collapse and a chemically inert barrier against degradation, the sandwich architecture successfully stabilizes metallenes.
Our investigation of stability mechanisms suggests that the specific mechanical properties, lattice constants, and larger interaction between metals and the sandwich layer of TMDs provide a more robust environment for structural confinement, allowing a broader range of metals to maintain their 2D integrity regardless of their natural coordination preferences. 
Also, structural confinement is supported by keeping the interfacial lattice mismatch below $3$\%, thereby preventing internal stresses that would otherwise cause the metal to lose its planarity. Simultaneously, the atomically flat, dangling-bond-free surfaces of the 2D encapsulation layers prevent the metal from forming strong, property-altering bonds with the substrate. 
Ultimately, these insights provide clear, quantitative guidance for selecting optimal heterostructure configurations, offering a practical roadmap for researchers to synthesize and utilize metallenes in real-world applications.

\section*{Data availability}
The initial structure trajectories used for the high-throughput calculations, ML-D3 optimized trajectories, alongside a database containing all the structurally and dynamically stable metallene sandwich heterostructures with other relevant properties, have been deposited at \href{https://doi.org/10.6084/m9.figshare.32724714}{https://doi.org/10.6084/m9.figshare.32724714}.

\section*{Acknowledgments}
We acknowledge the Jane and Aatos Erkko Foundation for funding (project EcoMet). We also thank CSC—IT Center for Science and the EuroHPC Joint Undertaking for awarding this project access to the EuroHPC supercomputer LUMI, hosted by CSC (Finland) and the LUMI consortium through a EuroHPC Regular Access call for computational resources.

\section*{Supporting information}
The following file is available free of charge.
\begin{itemize}
  \item Supporting Information: Computational methods and simulation parameters, additional details on designing metallene sandwich heterostructures, additional data of benchmarks, stable structures phonon dispersions, and sample Python scripts for working with supplementary data. (PDF)
\end{itemize}

\printbibliography

\end{document}